\documentclass[twocolumn,showpacs,aps,prl,superscriptaddress]{revtex4}

\usepackage{graphicx}
\usepackage{dcolumn}
\usepackage{amsmath}
\usepackage{epsfig}

\RequirePackage{xspace}

\usepackage{relsize}
\def\babar{\mbox{\slshape B\kern-0.1em{\smaller A}\kern-0.1em
    B\kern-0.1em{\smaller A\kern-0.2em R}}}

\def\epem       {\ensuremath{e^+e^-}\xspace}

\def\q     {\ensuremath{q}\xspace}

\def\qqbar {\ensuremath{q\overline q}\xspace}
\def\u     {\ensuremath{u}\xspace}
\def\ubar  {\ensuremath{\overline u}\xspace}

\def\d     {\ensuremath{d}\xspace}

\def\s     {\ensuremath{s}\xspace}

\def\c     {\ensuremath{c}\xspace}

\def\bbar  {\ensuremath{\overline b}\xspace}

\def\piz   {\ensuremath{\pi^0}\xspace}

\def\Kbar  {\kern 0.2em\overline{\kern -0.2em K}{}\xspace}

\def\Kz    {\ensuremath{K^0}\xspace}
\def\Kzb   {\ensuremath{\Kbar^0}\xspace}
\def\KzKzb {\ensuremath{\Kz \kern -0.16em \Kzb}\xspace}
\def\Kp    {\ensuremath{K^+}\xspace}
\def\Km    {\ensuremath{K^-}\xspace}

\def\KpKm  {\ensuremath{\Kp \kern -0.16em \Km}\xspace}

\def\Dbar    {\kern 0.2em\overline{\kern -0.2em D}{}\xspace}

\def\Dz      {\ensuremath{D^0}\xspace}
\def\Dzb     {\ensuremath{\Dbar^0}\xspace}
\def\DzDzb   {\ensuremath{\Dz {\kern -0.16em \Dzb}}\xspace}
\def\Dp      {\ensuremath{D^+}\xspace}
\def\Dm      {\ensuremath{D^-}\xspace}

\def\DpDm    {\ensuremath{\Dp {\kern -0.16em \Dm}}\xspace}

\def\B       {\ensuremath{B}\xspace}
\def\Bbar    {\kern 0.18em\overline{\kern -0.18em B}{}\xspace}

\def\BB      {\ensuremath{B\Bbar}\xspace} 
\def\Bz      {\ensuremath{B^0}\xspace}
\def\Bzb     {\ensuremath{\Bbar^0}\xspace}
\def\BzBzb   {\ensuremath{\Bz {\kern -0.16em \Bzb}}\xspace}
\def\Bu      {\ensuremath{B^+}\xspace}
\def\Bub     {\ensuremath{B^-}\xspace}

\def\Bpm     {\ensuremath{B^\pm}\xspace}

\def\BpBm    {\ensuremath{\Bu {\kern -0.16em \Bub}}\xspace}

\def\BorBbar    {\kern 0.18em\optbar{\kern -0.18em B}{}\xspace}
\def\DorDbar    {\kern 0.18em\optbar{\kern -0.18em D}{}\xspace}
\def\KorKbar    {\kern 0.18em\optbar{\kern -0.18em K}{}\xspace}

\mathchardef\Upsilon="7107
\def\Y#1S{\ensuremath{\Upsilon{(#1S)}}\xspace}

\mathchardef\Deltares="7101
\mathchardef\Xi="7104
\mathchardef\Lambda="7103
\mathchardef\Sigma="7106
\mathchardef\Omega="710A

\def\Deltabar{\kern 0.25em\overline{\kern -0.25em \Deltares}{}\xspace}
\def\Lbar{\kern 0.2em\overline{\kern -0.2em\Lambda\kern 0.05em}\kern-0.05em{}\xspace}
\def\Sigbar{\kern 0.2em\overline{\kern -0.2em \Sigma}{}\xspace}
\def\Xibar{\kern 0.2em\overline{\kern -0.2em \Xi}{}\xspace}
\def\Obar{\kern 0.2em\overline{\kern -0.2em \Omega}{}\xspace}
\def\Nbar{\kern 0.2em\overline{\kern -0.2em N}{}\xspace}
\def\Xb{\kern 0.2em\overline{\kern -0.2em X}{}\xspace}

\def\BR         {{\ensuremath{\cal B}\xspace}}

\def\mes        {\mbox{$m_{\rm ES}$}\xspace}

\def\DeltaE     {\mbox{$\Delta E$}\xspace}

\newcommand{\tev}{\ensuremath{\mathrm{\,Te\kern -0.1em V}}\xspace}
\newcommand{\gev}{\ensuremath{\mathrm{\,Ge\kern -0.1em V}}\xspace}
\newcommand{\mev}{\ensuremath{\mathrm{\,Me\kern -0.1em V}}\xspace}
\newcommand{\kev}{\ensuremath{\mathrm{\,ke\kern -0.1em V}}\xspace}
\newcommand{\ev}{\ensuremath{\mathrm{\,e\kern -0.1em V}}\xspace}
\newcommand{\gevc}{\ensuremath{{\mathrm{\,Ge\kern -0.1em V\!/}c}}\xspace}
\newcommand{\mevc}{\ensuremath{{\mathrm{\,Me\kern -0.1em V\!/}c}}\xspace}
\newcommand{\gevcc}{\ensuremath{{\mathrm{\,Ge\kern -0.1em V\!/}c^2}}\xspace}
\newcommand{\mevcc}{\ensuremath{{\mathrm{\,Me\kern -0.1em V\!/}c^2}}\xspace}

\def\m    {\ensuremath{{\rm \,m}}\xspace}

\def\mus  {\ensuremath{\rm \,\mus}\xspace}

\def\mus        {\ensuremath{\,\mu{\rm s}}\xspace}    

\def\to                 {\ensuremath{\rightarrow}\xspace}

\newcommand{\stat}{\ensuremath{\mathrm{(stat)}}\xspace}
\newcommand{\syst}{\ensuremath{\mathrm{(syst)}}\xspace}

\def\pep2{PEP-II}

\def\gsim{{~\raise.15em\hbox{$>$}\kern-.85em
          \lower.35em\hbox{$\sim$}~}\xspace}
\def\lsim{{~\raise.15em\hbox{$<$}\kern-.85em
          \lower.35em\hbox{$\sim$}~}\xspace}

\def\CP                {\ensuremath{C\!P}\xspace}

\xspace

\newcommand{\jprlBase}       {Phys.\ Rev.\ Lett.\xspace}
\newcommand{\jprBase}        {Phys.\ Rev.\xspace}
\newcommand{\jplBase}        {Phys.\ Lett.\xspace}
\newcommand{\nimBaseA}       {Nucl.\ Instr.\ Methods Phys.\ Res., Sect.\ A\xspace}

\newcommand{\zpBase}         {Z.\ Phys.\xspace}

\newcommand{\cpc}       [1]  {{Comput.\ Phys.\ Commun.\ {\bf #1}}}

\newcommand{\nima}      [1]  {\nimBaseA~{\bf #1}}

\newcommand{\plb}       [1]  {\jplBase\ B~{\bf #1}}

\newcommand{\jprl}      [1]  {\jprlBase\ {\bf #1}}
\newcommand{\jprd}      [1]  {\jprBase\ D~{\bf #1}}

\newcommand{\progtp}    [1]  {{Prog.\ Theor.\ Phys.\ {\bf #1}}}

\newcommand{\zpc}       [1]  {\zpBase\ C~{\bf #1}}

\def\jetset74   {\mbox{\tt Jetset \hspace{-0.5em}7.\hspace{-0.2em}4}\xspace}

\def\ptrue             {\ensuremath{f_{L}}}

\def\maoneb            {\ensuremath{m_{a^+_1}}}
\def\fish              {\ensuremath{\cal{F}}}
\def\calB              {\ensuremath{{\BR}}}

\def\fixed             {\ensuremath{({\rm fixed})}} 
\def\nominalcontnb     {\ensuremath{25798 \pm 182}}
\def\nominalsignb      {\ensuremath{90 \pm 38 \stat}}
\def\signalyield       {\ensuremath{68 \pm 38 \stat}}
\def\signalyieldb      {\ensuremath{68 \pm 38 \stat ^{+45}_{-56} \syst}}
\def\nominalsignbprime {\ensuremath{42 \pm 98}}

\def\nominalbr         {\ensuremath{(15.7 \pm 8.7 \stat ^{+10.3}_{-12.8} \syst)\e{-6}}}

\def\nominalul         {\ensuremath{30\e{-6} \, (90\%\ {\rm C.L.})}}
\def\doublenominalul   {\ensuremath{61\e{-6} \, (90\%\ {\rm C.L.})}}
\def\theory            {\ensuremath{43\e{-6}}}

\def\significancesystb {\ensuremath{0.95\sigma}}

\def\aone              {\ensuremath {a_1^+}}
\def\signal            {\ensuremath {\Bz\rightarrow a_1^\pm \rho^\mp }}
\def\signalb           {\ensuremath {\Bz\rightarrow a_1^+ \rho^- }}

\def\rhoz              {\ensuremath {\rho^0}}
\def\Bztoacrhoc        {\ensuremath{\Bz \to a_1^\pm \rho^\mp}\xspace}

\def\btodstarrho {\ensuremath{\B \to {D}^{\star}\:\rho}}

\newcommand\vud {\ensuremath{V_{ud}}}

\newcommand\vub {\ensuremath{V_{ub}}}

\newcommand\vcb {\ensuremath{V_{cb}}}
\newcommand\vtd {\ensuremath{V_{td}}}

\newcommand\vtb {\ensuremath{V_{tb}}}

\newcommand\sgline{\noalign{\vskip 0.10truecm\hrule\vskip 0.10truecm}}
\newcommand\sgdline{\noalign{\vskip 0.10truecm\hrule\vskip 0.05truecm\hrule\vskip 0.10truecm}}

\newcommand{\e}      [1]   { {\ensuremath{ \times 10^{ {#1} } }}}

\newcommand{\BABARPubYear}    {06}
\newcommand{\BABARPubNumber}  {017}

\newcommand{\SLACPubNumber} {11850}

\def\figurebox#1#2#3{%
    \def\arg{#3}%
    \ifx\arg\empty
    {\hfill\vbox{\hsize#2\hrule\hbox to #2{\vrule\hfill\vbox to #1{\hsize#2\vfill}\vrule}\hrule}\hfill}%
    \else
    {\hfill\epsfbox{#3}\hfill}%
    \fi}

\begin{document}

\preprint{\babar-PUB-\BABARPubYear/\BABARPubNumber} 
\preprint{SLAC-PUB-\SLACPubNumber} 

\begin{flushleft}
\babar-PUB-\BABARPubYear/\BABARPubNumber\\
SLAC-PUB-\SLACPubNumber\\[10mm]
\end{flushleft}
\vspace{-1.0cm}

\title{
{\large \bf
Search for the decay {\boldmath\Bztoacrhoc}} 
}

%
\author{B.~Aubert}
\author{R.~Barate}
\author{M.~Bona}
\author{D.~Boutigny}
\author{F.~Couderc}
\author{Y.~Karyotakis}
\author{J.~P.~Lees}
\author{V.~Poireau}
\author{V.~Tisserand}
\author{A.~Zghiche}
\affiliation{Laboratoire de Physique des Particules, F-74941 Annecy-le-Vieux, France }
% ...
\collaboration{The \babar\ Collaboration}
\noaffiliation

\date{\today}

\begin{abstract}
We present a search for the rare \B-meson decay \Bztoacrhoc with
 $a_1^\pm\rightarrow\pi^+\pi^-\pi^\pm$. We use
 (110 $\pm$ 1.2) $\times$ 10$^6$ $\Y4S\rightarrow\BB$ decays collected
 with the \babar\ detector at the \pep2\ asymmetric-energy \B\ Factory at SLAC.
We obtain an upper limit of \nominalul\ for the branching fraction product
$\calB(\Bz\rightarrow{a_1^\pm}\rho^\mp)\calB({a_1^\pm}\rightarrow\pi^+\pi^-\pi^\pm)$,
where we assume that the $a_1^\pm$ decays exclusively to $\rhoz\pi^\pm$.

\end{abstract}


\maketitle

In the Standard Model, \CP-violating effects in the \B-meson 
system arise from a single phase in the 
Cabibbo-Kobayashi-Maskawa (CKM) quark-mixing matrix~\cite{CKM}.
The decay \signal\ proceeds via a $\bbar \to \u\ubar \d$ transition~\cite{charge}, and
interference between direct decay and
decay after $\Bz\Bzb$ mixing results 
in a time-dependent decay-rate asymmetry that is sensitive 
to the angle $\alpha\equiv {\mathrm{arg}} [-\vtd\vtb^*/\vud\vub^*]$~\cite{ref:alphareview} 
in the unitarity triangle of the CKM matrix. 
An additional motivation for studying \signal\ is that this is a significant 
background to $\B \to \rho\rho$ decays, e.g.~\cite{ref:rhoprhom,ref:rhoprhoz,ref:rhozrhoz,ref:bellerhoprhom,ref:bellerhoprhoz},
which currently provide the most accurate measurement of $\alpha$.
The ARGUS experiment previously searched for the decay \signal, which resulted in an upper limit of 
$\calB(\Bz \to a_1^\pm \rho^\mp) < 3.4 \times 10^{-3}$ (90\% C.L.)~\cite{ref:argusa1rho}.
This paper presents the result of a search for \signal\ with $a_1^\pm\to \pi^+\pi^-\pi^\pm$,
where we assume that the $a_1^\pm$ decays exclusively to $\rhoz\pi^\pm$. 
A theoretical prediction of the branching fraction $\calB(\Bz\rightarrow{a_1^\pm}\rho^\mp)\calB({a_1^\pm}\rightarrow (3\pi)^\pm)$ 
has been made by Bauer, Stech and Wirbel~\cite{ref:zpc} within the framework of the
factorization model. They predict a value of $\theory$, assuming $|\vub/\vcb|$ = 0.08.

The data used in this analysis were collected with the \babar\ detector at the 
\pep2\ asymmetric-energy \B\ Factory at SLAC during the years 2003-2004. This 
represents a total integrated luminosity of 100 fb$^{-1}$ taken at the 
$\Upsilon$(4S) resonance (on-peak), corresponding to a sample of 110 $\pm$ 1.2 
million $\BB$ pairs. An additional 21.6 fb$^{-1}$ of data, collected 
at approximately 40 \mev\ below the $\Upsilon$(4S) resonance (off-peak), were used to study
background from \epem\to\qqbar ($\q = \u,\d,\s,\c$) continuum events.

The \babar\ detector is described in detail elsewhere~\cite{ref:babar}. 
Surrounding the interaction point is a silicon vertex tracker (SVT) with 5 double-sided layers
which measures the impact parameters of charged particle tracks in both the plane transverse
to, and along the beam direction. A 40 layer drift chamber (DCH) surrounds the SVT and provides measurements
of the transverse momenta for charged particles. Both the SVT and the DCH
 operate in the magnetic field of a 1.5 T solenoid. Charged hadron 
identification is achieved through measurements of particle energy-loss 
 in the tracking system and the Cherenkov angle obtained 
from a detector of internally reflected Cherenkov light.  
A CsI(Tl) electromagnetic calorimeter (EMC) provides photon detection, electron 
identification, and $\piz$ reconstruction.
Finally, the instrumented flux return of the magnet allows discrimination of muons from pions.

We reconstruct \signalb\ candidates from combinations of $a_1^+\rightarrow\pi^+\pi^-\pi^+$ and
$\rho^-\rightarrow\piz\pi^-$ candidates. The $a_1(1260)\to 3\pi$ decay proceeds mainly through 
the intermediate states $(\pi\pi)_\rho \pi$ and $(\pi\pi)_\sigma \pi$~\cite{ref:cleo}.  We 
do not distinguish between the dominant P-wave $(\pi\pi)_\rho$ and S-wave 
$(\pi\pi)_\sigma$ in the channel $\pi^+\pi^-$. The Monte Carlo (MC) signal events are simulated
as \Bz\ decays to $a_1^+(1260)\rho^-$ with $a_1^+ \to \rho^0\pi^+$ using the GEANT4-based~\cite{ref:geant} 
\babar\ MC simulation.
Possible contributions from \Bz\ decays to $a_2^+(1320)\rho^-$ and $\pi^+(1300)\rho^-$ are investigated. 

We only consider events that have a minimum of one 
\piz\ and four charged tracks, where the charged tracks are required to be inconsistent with 
lepton, proton and kaon hypotheses.

We form $\piz\rightarrow\gamma\gamma$ candidates from pairs of
photon candidates that have been identified as localized energy deposits in the EMC 
that have the lateral energy distribution expected for a photon.
Each photon is required to have an energy $E_\gamma > 50 \mev$, and the $\piz$ 
is required to have an invariant mass of $0.10 < m_{\gamma\gamma} < 0.16 \gevcc$.

The $\rho^-$ mesons are formed from one track that is consistent 
with a $\pi^-$ and the aforementioned $\piz$ candidate.  The candidate $\rho^-$ 
 is required to have an invariant mass of $0.5 < m_{\rho^-} < 1.1 \gevcc$.   
We also constrain the cosine of the angle between the \piz\ momentum and the direction
opposite to the \Bz\ in the $\rho^-$ rest frame ($\cos\theta_{\rho^-}$) to be between $-$0.9 and 
0.98. This removes backgrounds which peak at the extremes of the distribution where the 
signal reconstruction efficiency also falls off.

We form the $a_1^+$ candidate from combinations of three charged pions. 
We first form a $\rho^{0} \rightarrow \pi^+\pi^-$ candidate from two oppositely charged tracks. This combination
is required to have an invariant mass of $0.4 < m_{\rhoz} < 1.1 \gevcc$. The $a_1^+$ candidate is then formed
by adding another charged track to the \rhoz, and requiring that the mass of the 
\aone\ satisfies $0.6 < \maoneb < 1.5 \gevcc$. The vertex of the \B-candidate is 
constrained to originate from the beam spot. In order to reduce background from continuum events we require 
that $|\cos(\theta_{T})| < 0.7$, where $\theta_{T}$ is the angle between the \B\ thrust axis and that of 
the rest of the event (ROE).

We use two kinematic variables, $\mes$ and $\DeltaE$, in order to isolate any signal.
We define the beam-energy substituted mass $\mes=\sqrt{(\sqrt{s}/2)^2-(p_B^{\rm *})^2}$, 
where $\sqrt{s}$ is the \epem\ center-of-mass (CM) energy.
The second kinematic variable, \DeltaE, is the difference between the \B-candidate energy and the
beam energy in the CM frame. We require $\mes > 5.25 \gevcc$ and $-0.15 < \DeltaE < 0.1 \gev$.

Additional separation between signal and continuum is obtained by combining several kinematic and 
topological variables into a Fisher discriminant ($\fish$)~\cite{ref:fisher}. 
The variables $L_0$, $L_2$, and $|\cos \theta_{TR}|$, and the output of a multivariate tagging 
algorithm~\cite{ref:tag} are used as inputs to $\fish$. $L_0$ and $L_2$ are defined as 
\begin{eqnarray}
L_0 = \sum_{\mathrm{ROE}} |p_i^*|, \
L_2 = \sum_{\mathrm{ROE}} |p_i^*| \cos(\theta_i)^2,
\end{eqnarray}
where the sum is over the ROE, $p_i^*$ is the particle momentum in the CM frame. $\theta_i$ is the 
angle of the particle direction relative to the thrust axis of the \B-candidate, and $\cos \theta_{TR}$
is the cosine of the angle between the \B thrust axis and the beam axis. 
The multivariate tagging algorithm identifies the flavor of the other \B in the event to be either 
a \Bz or \Bzb. The output of this algorithm is ranked into categories of different signal purity.

We expect the polarization of the $\aone\rho^-$ final state to be predominantly longitudinal,
as was found in the similar decay $\B \to\rho\rho$~\cite{ref:rhoprhom,ref:rhoprhoz,ref:rhozrhoz,ref:bellerhoprhom,ref:bellerhoprhoz}.
We have used both longitudinal and transverse polarized signal MC simulated data in 
this analysis. After applying the selection cuts above, we have 2.8 (2.3) longitudinal (transverse) polarized signal 
MC simulated data candidates per event. 

We define as self-cross-feed (SCF) the set of candidates that were incorrectly reconstructed
from particles in events that contain a true signal candidate. We select one \B\ candidate per
event in which the mass of the reconstructed $\rhoz$ is closest to that of the true $\rhoz$
mass~\cite{ref:pdg}. Choosing the candidate using the \rhoz\ mass reduces
the SCF fraction by 18\% relative to a random selection. To avoid potentially biasing our 
final result, we do not use information from the \rhoz\ 
meson in the remainder of the analysis. After all selection cuts have been applied, the 
longitudinal and transverse signal SCF fractions are 0.58 and 0.42, respectively. The selection 
efficiency of longitudinal (transverse) signal is 9.44\% (10.15\%).  

Besides the continuum background we also have background from \B\ decays. 
We divide the \B-background into the
following four categories according to \B-meson charge and the charm
content of the final states: (i) $\Bz\rightarrow$ charm, (ii) $\Bz\rightarrow$ charmless, 
(iii) $\Bpm\rightarrow$ charm and (iv) $\Bpm\rightarrow$ charmless. From large samples of 
inclusive MC simulated data we expect 
2394, 424, 3281 and 215 events of these background types, respectively. In addition, a number of 
exclusive \B-background modes that have a similar final state to the signal were studied. This
includes those that have an intermediate $a_1$ meson in the decay. None of these modes were seen to have a
significant efficiency after the selection cuts had been applied.  

We perform an extended unbinned maximum likelihood fit to the data.  The likelihood
model has the following types: (i)-(iv) the four aforementioned inclusive \B-background categories, 
(v) true signal, (vi) SCF signal and the (vii) \epem\to\qqbar ($\q = \u,\d,\s,\c$) continuum background.
The probability density function (PDF) for each event $i$ has the form
$P_{i,c} = P_{i,c}(\mes, \DeltaE, {\fish}, \maoneb, m_{\rho^-}, \cos\theta_{\rho^-} )$.
From these individual PDFs the total likelihood
\begin{equation}
 \mathrm{L} =   e^{-n^{\prime}}\prod_{i=1}^{n} \sum_{c} N_{c} P_{i,c}, \label{eqn:likelihood}
\end{equation}
is constructed. The parameters $n$ and $n^{\prime}$ are the numbers of selected on-peak events, and the
sum of the yields $N_{c}$, where $c$ is one of the seven types in the likelihood model.
Correlations between the $m_{\rho^-}$ and $\cos\theta_{\rho^-}$ variables are taken
into account for the real (T) and fake combinatorial (F) $\rho^-$ candidates.
All other correlations between fit variables are seen to be small. However, the effect of ignoring
them results in a bias on the fitted signal yield which is discussed below. 

Each of the signal distributions has a signal yield and a polarization fraction that
 are denoted by $N_{sig}$ ($N_{sig}^\prime$) and \ptrue\ ($\ptrue^\prime$), for the
true (SCF) signal, such that the sum of the true and SCF signal is described by
\begin{eqnarray}
N_{sig}[\ptrue P_i^{long, true} + (1 - \ptrue) P_i^{tran, true} ] + \nonumber \\
N^\prime_{sig}[ \ptrue^\prime P_i^{long, SCF} + (1 - \ptrue^\prime) P_i^{tran,\ SCF}].
\end{eqnarray}
The continuum yield, $N_{sig}$, $N_{sig}^\prime$, and the parameters of the continuum \mes\ and \DeltaE\ PDFs
are allowed to vary in the fit.
Under the assumption that no significant signal is observed, the value of $\ptrue$ is fixed to 1.0 
in the fit. The value of $\ptrue^\prime$ is also fixed to 1.0 in the fit since it is highly correlated with 
$N_{sig}$, $N_{sig}^\prime$ and \ptrue. 
Only the fitted value of $N_{sig}$ is used to derive the final result.
We also fix the \B-background yields to the aforementioned values. 
\begin{table*}[!t]
\caption{The types of PDFs used to model the different variables for each component in the likelihood fit,
         where the PDFs underlined have their parameters varying in the nominal fit. 
         The abbreviations are: G = Gaussian, G2 = Double Gaussian, G3 = Triple Gaussian, 
         CB = Crystal Ball (a Gaussian with a low side exponential tail)~\cite{ref:cb}, 
         ARGUS = ARGUS function $x\sqrt{1-x^2}\exp{\left[-\xi(1-x^2)\right]}$, with
         $x\equiv2\mes/\sqrt{s}$ and parameter $\xi$ \cite{ref:argus}, which is allowed to vary in the fit, 
         Pn = Polynomial of order n, BW = Breit-Wigner,
         helicity = $\cos^2\theta_{\rho^-}$ or  $\sin^2\theta_{\rho^-}$  depending on partial wave 
         which is modified by a quadratic acceptance function, BG m-hel = Background $\cos\theta_{\rho^-}$ 
         and $\m_\rho^-$ PDF of Eq.~\ref{eqn:bgmhel}, 
         off-peak = PDF taken from off-peak data, and 1D = smoothed 1D histogram.
}
\small
\begin{center}
\begin{tabular}{lcccccc} \sgline
Component                                  & $\mes$ & $\DeltaE$ & $\fish$ & $\maoneb$ & $\cos\theta_{\rho^-}$ & $m_{\rho^-}$ \\ \sgdline
signal (long/trans./true/SCF)              & CB     & CB+G  & G2 & G3 & helicity & BW+P4 \\
\qqbar                                     & $\underline{\mathrm{ARGUS}}$ & $\underline{\mathrm{P1}}$ & G2 (off-peak) & 1D (off-peak) & BG m-hel & BG m-hel\\
\Bz($\Bpm$)$\rightarrow$ charm (charmless) & NP & NP & NP & NP & BG m-hel & BG m-hel  \\\sgline
\end{tabular}
\end{center}
\label{tbl:pdfforms}
\end{table*}

The PDFs used for each component are given in 
Table \ref{tbl:pdfforms}.  The signal and \B-backgrounds are parameterized using MC. 
We use a non-parametric smoothing algorithm~\cite{ref:keys} when defining some 
of the background PDFs (as indicated by the abbreviation NP).
We account for the difference between F and T $\rho^-\rightarrow\pi^-\piz$ distributions in the background using the
product of 1D PDFs, denoted in Table~\ref{tbl:pdfforms} as `BG m-hel', such that
\begin{eqnarray}
P_{i} ( m_{\rho^-}, \cos\theta_{\rho^-}  ) &=& (1 - a_{T})P_{F, i} ( m_{\rho^-} ) P_{F, i} ( \cos\theta_{\rho^-}  ) + \nonumber \\
&& a_{T} P_{T, i} ( m_{\rho^-} ) P_{T, i} ( \cos\theta_{\rho^-}  ) ,\label{eqn:bgmhel}
\end{eqnarray}
where $a_{T}$ is the fraction of T events.
The continuum shape for $\cos\theta_{\rho^-}$ ($m_{\rho^-}$) is 
derived from off-peak (on-peak) data. The true $\rho^-$ resonance Breit Wigner shape uses $m_{\rho^-} = 0.77$ \gevcc, and 
$\Gamma = 0.150$ \gev~\cite{ref:pdg}. The parameterizations used for this PDF are summarized in Table~\ref{tbl:parambgmvhel}.
\begin{table}[!h]
\caption{The types of PDFs used to model the different background $\cos\theta_{\rho^-}$ and $m_{\rho^-}$ PDF shapes.
The abbreviations Pn and BW are defined in the caption of Table~\ref{tbl:pdfforms}.}
\small
\begin{center}
\begin{tabular}{lcccc} \sgline
Component                   & T $\cos\theta_{\rho^-}$ & T $m_{\rho^-}$ & F $\cos\theta_{\rho^-}$ & F $m_{\rho^-}$ \\ \sgdline
\qqbar                      & P2 & BW + P1 & P5 & P4 \\
$\Bpm\to$ charmless & P4 & BW      & P5 & P3 \\ 
\Bz$\to$ charmless  & P4 & BW      & P5 & P3 \\
$\Bpm\to$ charm     & P2 & BW      & P5 & P3 \\
\Bz$\to$ charm      & P2 & BW      & P5 & P3 \\\sgline 
\end{tabular}
\end{center}
\label{tbl:parambgmvhel}
\end{table}

\begin{figure*}[!ht]
\begin{center}
  \resizebox{18cm}{!}{
    \includegraphics{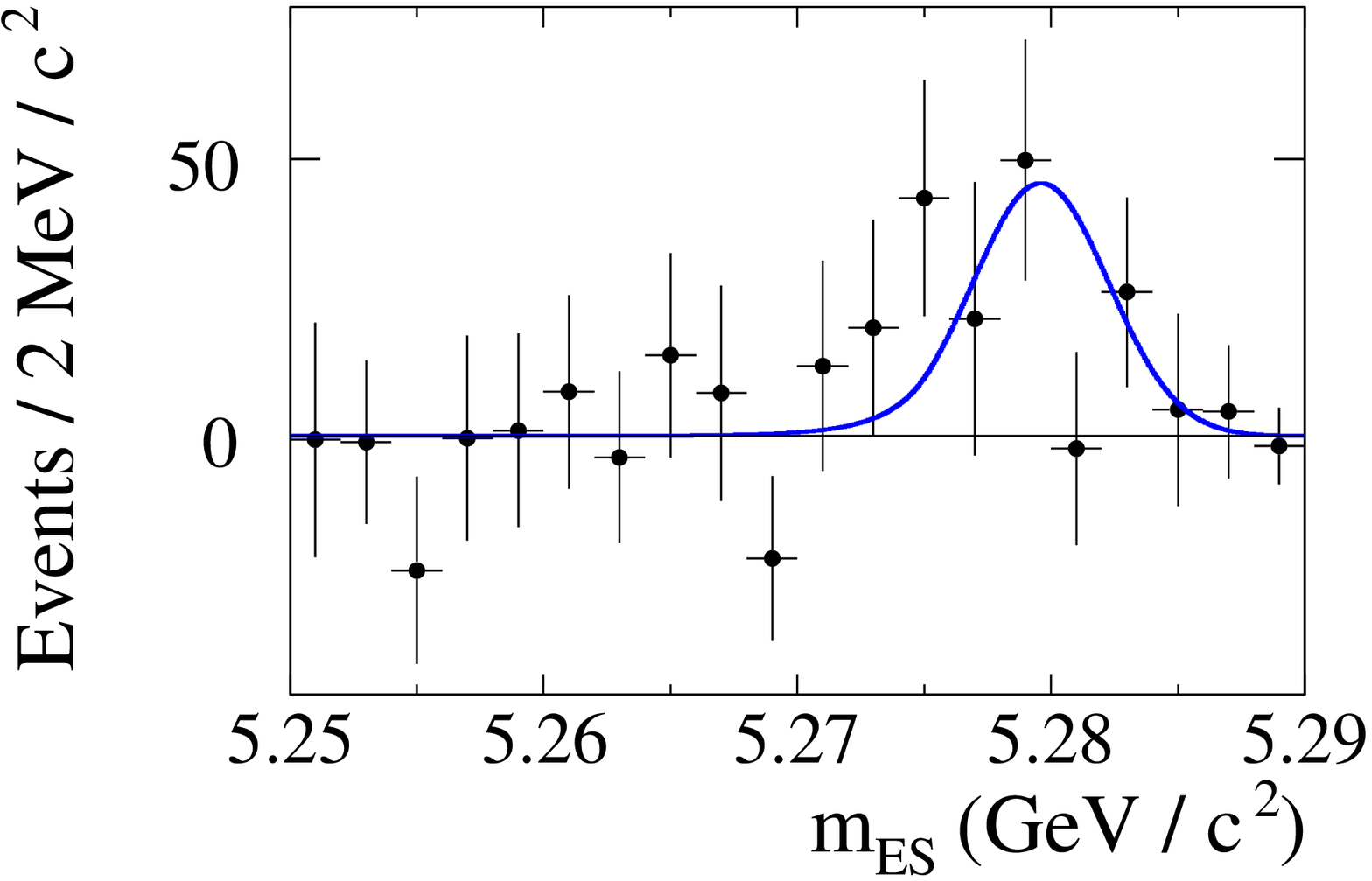}
    \includegraphics{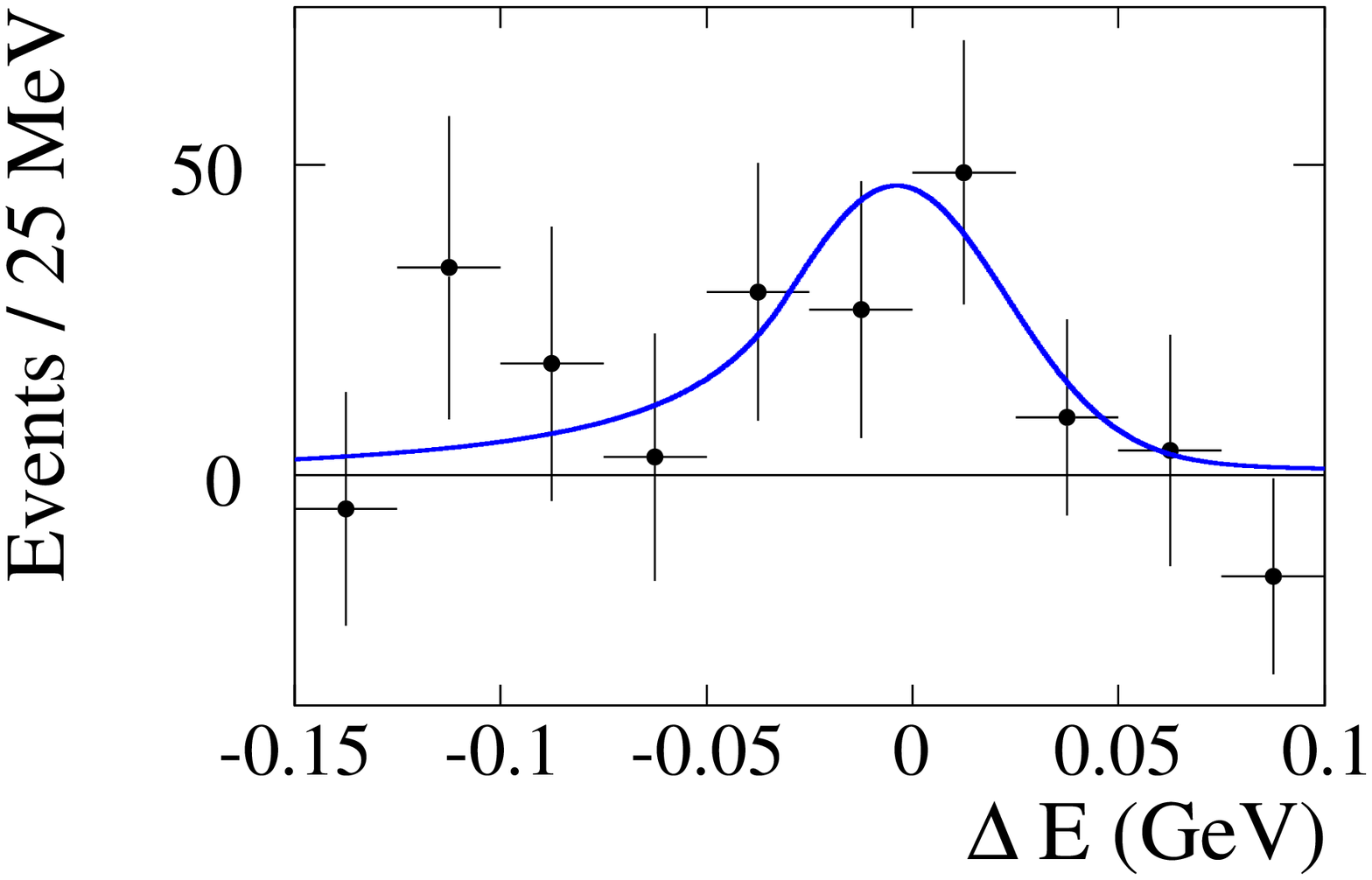}
    \includegraphics{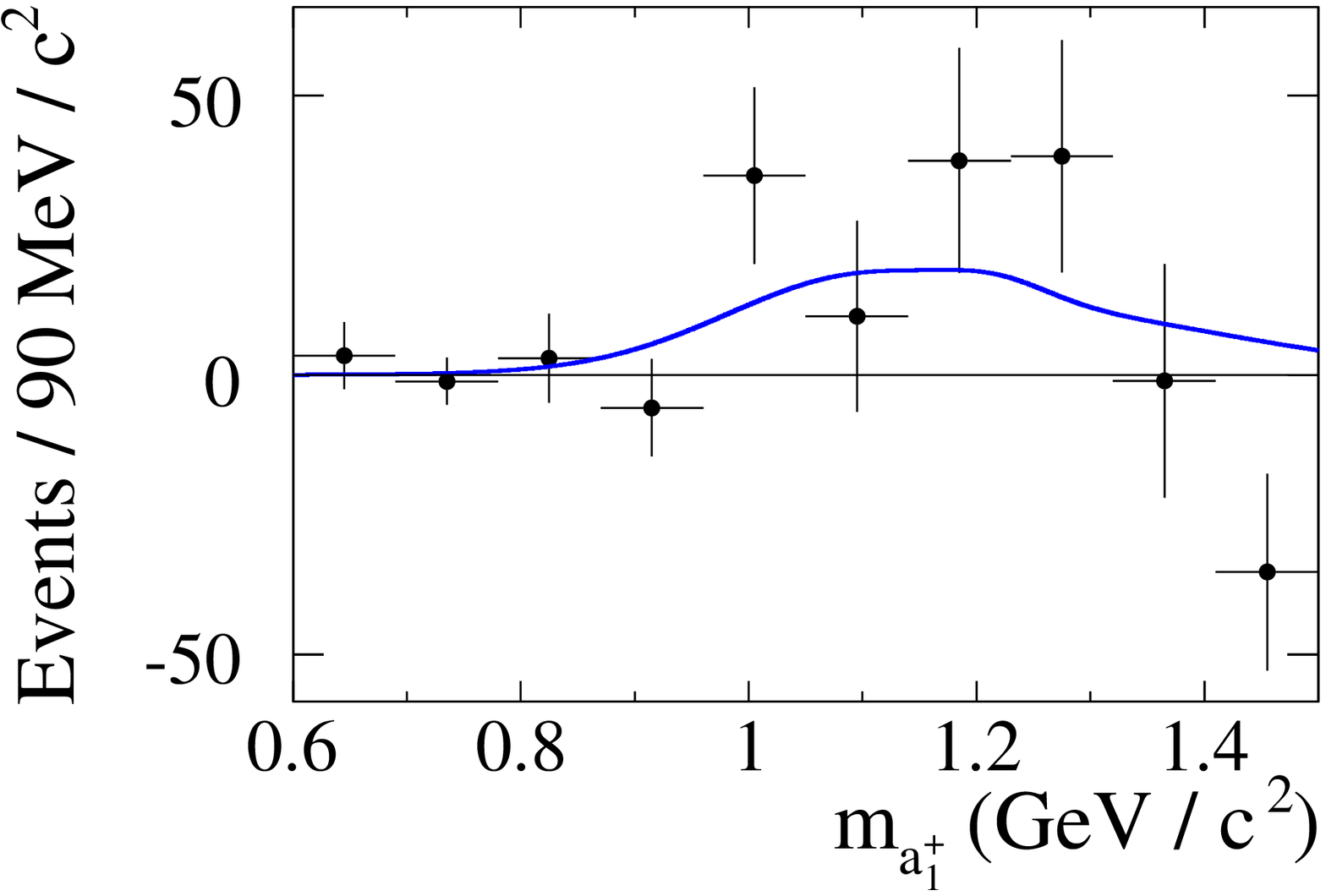}
  }
  \resizebox{18cm}{!}{
    \includegraphics{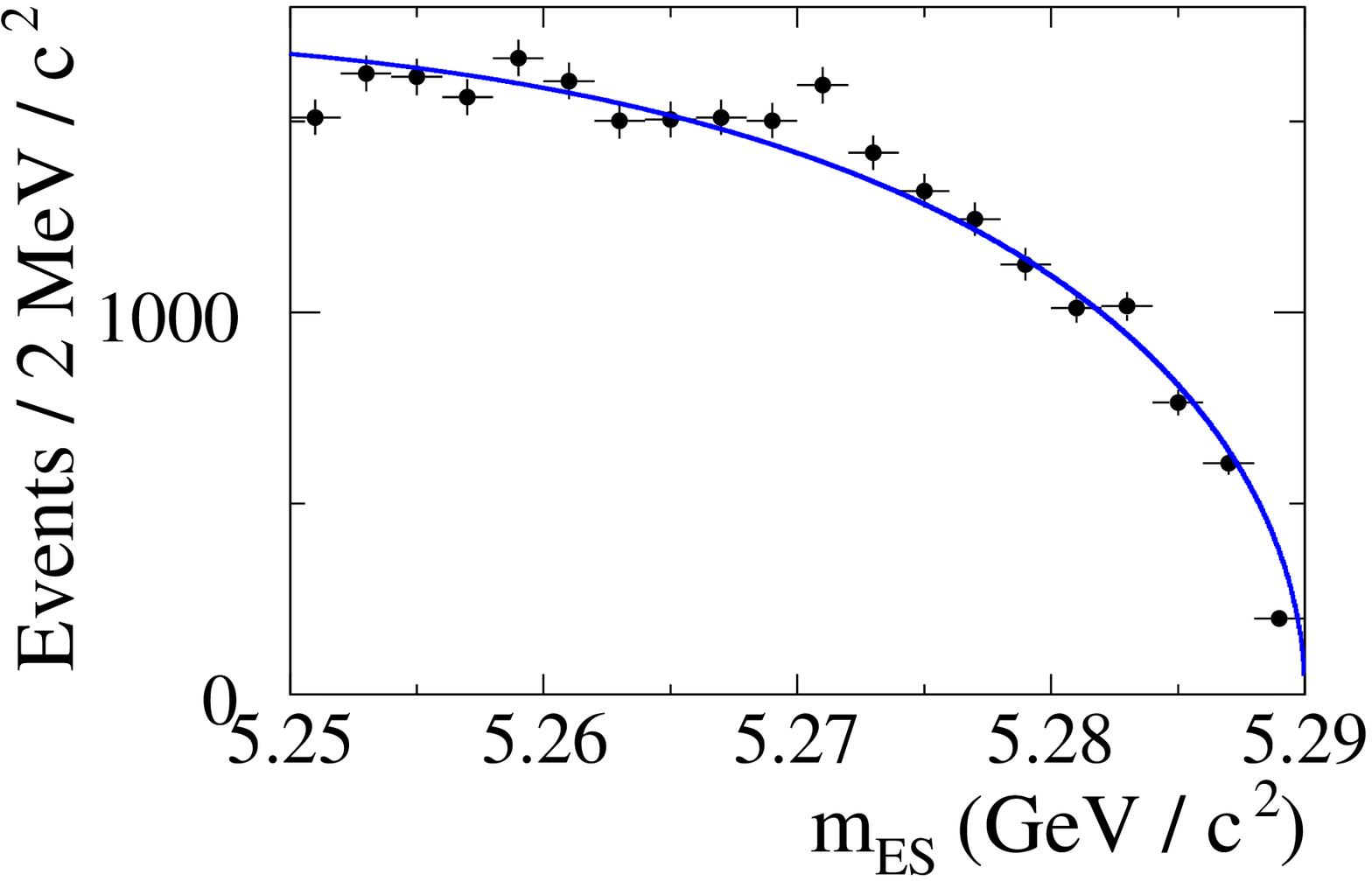}
    \includegraphics{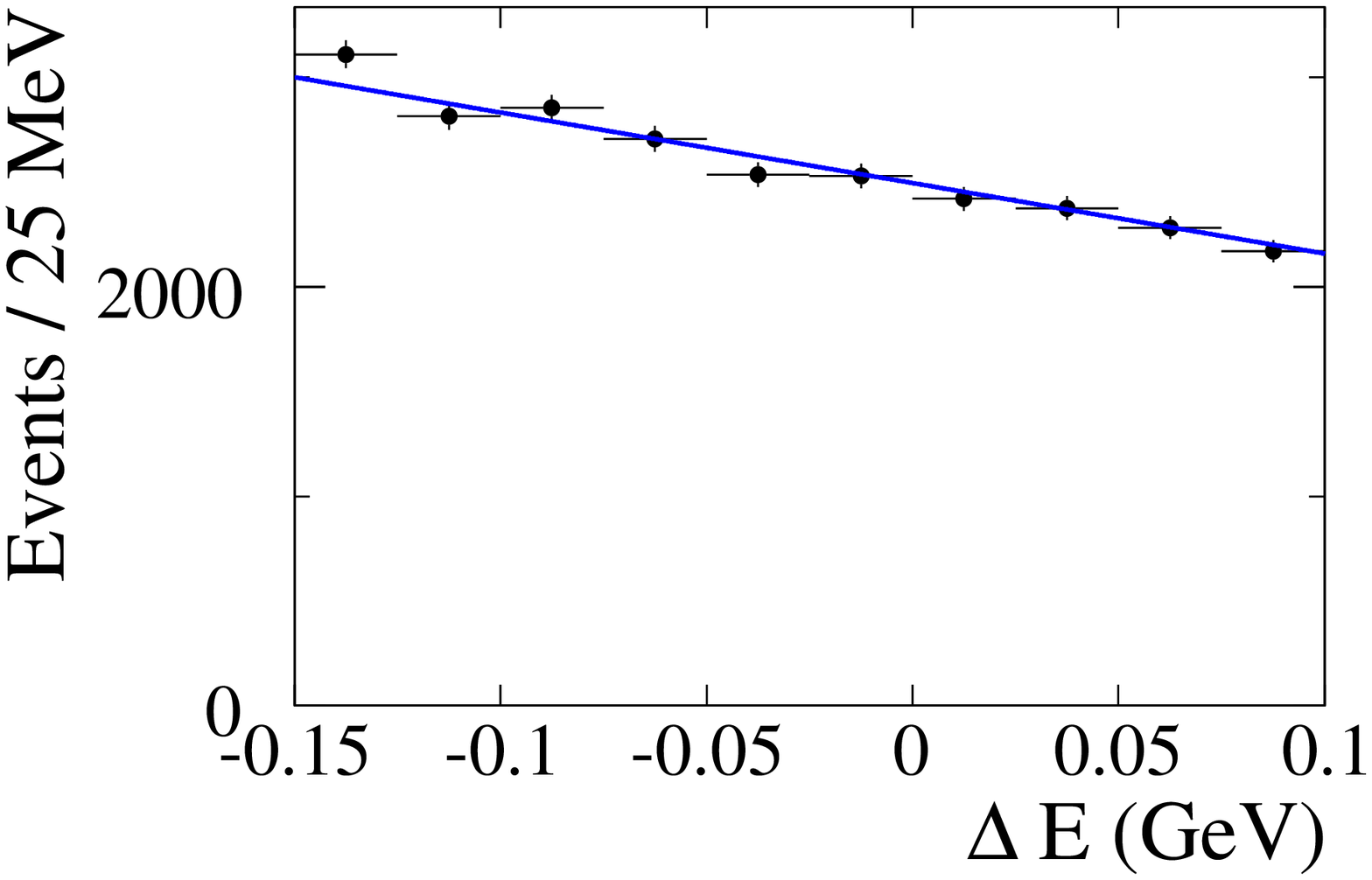}
    \includegraphics{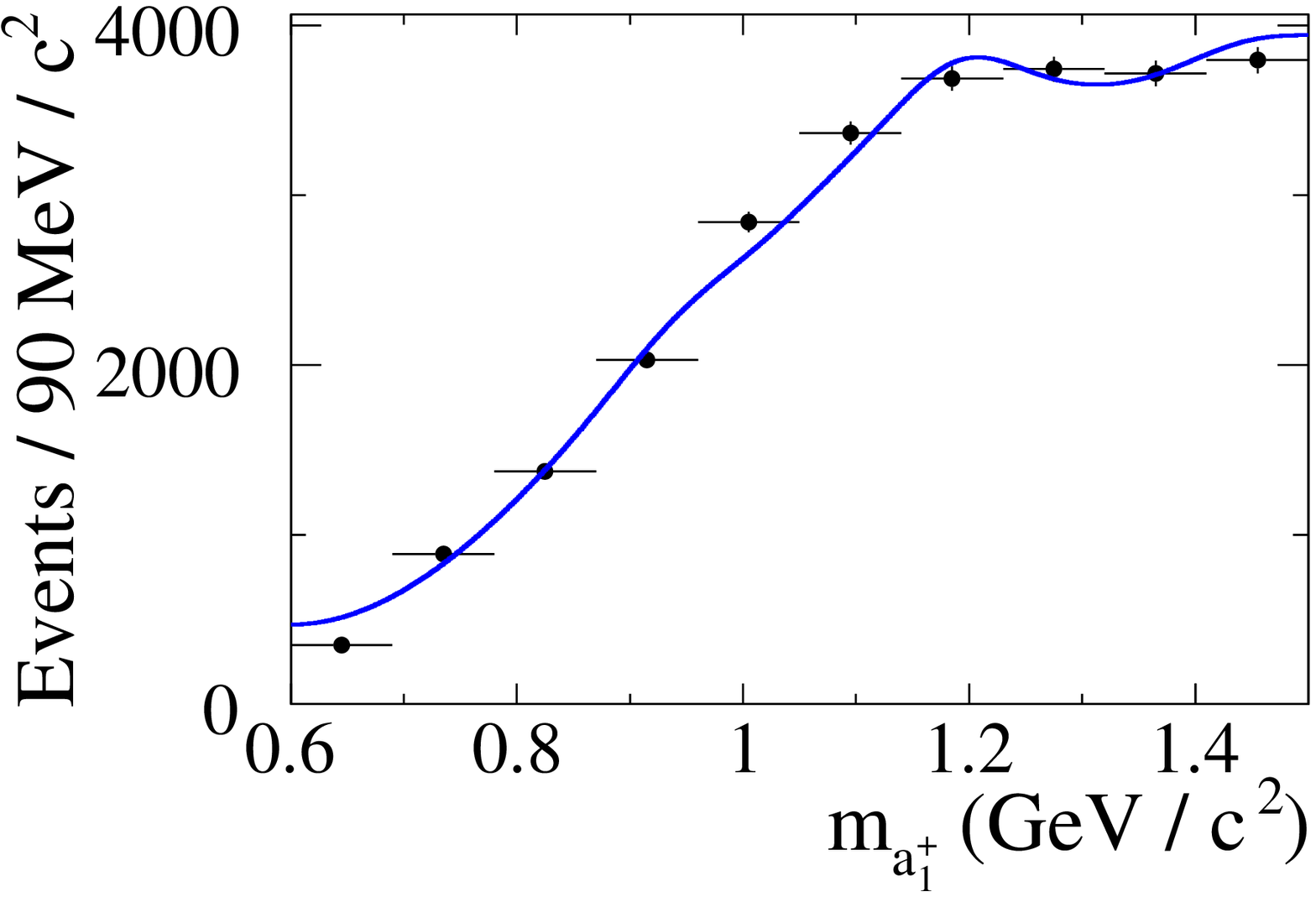}
  }
 \caption{The true signal (top) and continuum (bottom) distributions for (left to right) 
          $\mes$, $\DeltaE$, $\maoneb$ using the weighting technique described in 
          Ref.~\cite{ref:splots}.  The points represent the weighted data, and solid curves 
          represent the corresponding PDFs.\label{fig:splotsSig}} 
\end{center}
\end{figure*}

The results from the fit are $N_{sig} =$ \nominalsignb, 
$N_{sig}^\prime =$ \nominalsignbprime\ and a continuum yield of \nominalcontnb\ events.
The bias on the fitted signal yield is evaluated by performing ensembles of mock experiments using signal MC 
embedded into MC samples of background generated from the PDF. The bias is found to be $+$22 events (24$\%$),
resulting in a corrected signal yield of \signalyield. In Fig.~\ref{fig:splotsSig} we compare the true signal 
and continuum PDF shapes (solid curves) to the 
data (points) using the event-weighting technique described in Ref.~\cite{ref:splots}.  The distributions 
shown in Fig.~\ref{fig:splotsSig} are not corrected for fit bias, and the uncertainty on each of the data points is 
statistical.  No change in signal yield is seen when $a_2^+(1320)\rho^-$ and $\pi^+(1300)\rho^-$ components 
are included in the fit.

Table~\ref{tbl:additivesyst} summarises the systematic uncertainties on the signal yield.
Each entry in the table indicates one systematic effect, and the contributions are added in
quadrature to give the total presented. The uncertainty due to PDF
parameterisation is evaluated by variation of both the signal and the background PDF parameters
within their uncertainties about their nominal values. The uncertainty from the
continuum \mes\ and \DeltaE\ PDFs which are allowed to vary in the fit, are only included in
the quoted statistical uncertainty.
We assign a systematic uncertainty due to fit bias, evaluated as half of the fit bias 
correction on the signal yield.
To validate the expected \B-background yields, and to assign a systematic uncertainty we 
perform a number of cross-checks in which we allow the background yields to vary in turn 
when fitting the data. We use a control sample of $B\to D\rho$ events to determine the 
systematic uncertainty in the fraction of SCF signal events. The effect of exclusive \B\-meson 
decays to final states including $a_1$-mesons were evaluated using ensembles of mock 
experiments. In particular, the systematic uncertainty on the signal yield from 
neglecting $\B\to{a_1}{a_1}$ modes in the fit is 6 events. 
We assign a systematic uncertainty from using a relativistic Breit-Wigner 
with a Blatt-Weisskopf form factor with a range parameter of 3.0 $\gev^{-1}$ 
for the $a_1^+$ meson line shape. In the fit we assume that the $a_1^+$ meson width, 
$\Gamma_{a_1^+}$, is 400 \mev. We evaluate a systematic uncertainty due to this 
assumption by varying $\Gamma_{a_1^+}$ over the experimentally allowed range: 250 - 600\mev~\cite{ref:pdg}.
The difference in the distribution of \fish\
between data and MC is evaluated with a large sample of \btodstarrho\ decays.
\begin{table}[!h]
\caption{The systematic uncertainties on $N_{sig}$ (events).}
\begin{center}
\begin{tabular}{lc}\sgline
Source                                 & Uncertainty on $N_{sig}$ \\ \sgdline
PDF parameterisation                   & $\,^{+27}_{-30}$ \\
Fit bias                               & $\pm 11$\\
\B-background yields                   & $\,^{+29}_{-42}$ \\
SCF fraction                           & $\pm 7$ \\
Neglecting $\B\to a_1a_1$ modes in fit & $\pm 6$ \\ 
$a_1^+$ line shape                       & $\pm 10$ \\
$a_1^+$ width                            & $\pm 9$ \\
Fisher data/MC comparison              & $\pm 6$ \\\sgline 
Total                                  & $\,^{+45}_{-56}$ \\ \sgline
\end{tabular}
\end{center}
\label{tbl:additivesyst}
\end{table}
The systematic uncertainties that contribute to the branching fraction only through the efficiency come
from charged particle identification (6.0$\%$), $\piz$ meson reconstruction (3.0$\%$), 
tracking efficiency (3.2$\%$), and the number of \B\-meson pairs (1.1$\%$). 
The systematic error contribution from MC statistics is negligible.

\begin{figure}[!b]
\begin{center}
  \resizebox{8.5cm}{!}{
    \includegraphics{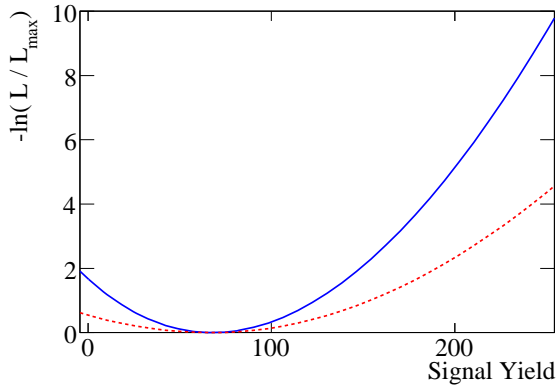}
  }
\caption{The $-\mathrm{ln}(\mathrm{L/L_{max}})$ distribution from the fit to data with \ptrue=1.0 \fixed.  
         This distribution has been corrected for fit bias.  The solid curve is for 
	 statistical errors only, and the dashed curve includes systematic errors.\label{fig:nllcurve}} 
\end{center}
\end{figure}

When the fit bias correction of $-$22 events is applied to the signal yield, and one 
accounts for systematic uncertainties, the significance of the result is 0.95 standard
deviations.
Figure~\ref{fig:nllcurve} shows the distribution of $-\mathrm{ln}(\mathrm{L/L_{max}})$ for 
the fit, with and without these systematic errors. $\mathrm{L_{max}}$ is the value of the 
likelihood corresponding to the nominal fit result.
 The branching fraction value for the fit-bias-corrected signal yield of $\signalyieldb$ is 
$\calB(\Bz\rightarrow{a_1^+}\rho^-)\calB({a_1^+}\rightarrow\pi^+\pi^-\pi^+)=\nominalbr$.  
This assumes that $\ptrue = 1.0$ and that the branching fraction of $a_1^+\to\pi^+\pi^-\pi^+$ = 0.5.  
As the signal yield obtained is not significant, we calculate the upper limit $x_\mathrm{UL}$,
by integrating the likelihood function (including systematic uncertainties) from 0 to $x_\mathrm{UL}$, 
for different physically allowed values of \ptrue, such that the C.L. of the upper limit is 90\%. As the signal efficiency 
is a function of \ptrue, we report the most conservative upper limit obtained, which corresponds to $\ptrue=1.0$. 
On doing this, an upper limit of \nominalul\ is obtained.

We have performed a search for the decay \signal\ in a data sample of 100 fb$^{-1}$.  
After correcting for fit bias and accounting for systematic uncertainties, 
the signal yield is \signalyieldb\ events, with a significance of \significancesystb. 
As there is no significant evidence for a signal, we place an upper limit of \nominalul\
on $\calB(\Bz\rightarrow{a_1^+}\rho^-)\calB({a_1^+}\rightarrow\pi^+\pi^-\pi^+)$,
where we assume that the $a_1^+$ decays exclusively to $\rhoz\pi^+$.
Assuming $\calB({a_1^+}\rightarrow\pi^+\pi^-\pi^+)$ is equal to $\calB({a_1^+}\rightarrow\pi^+\pi^0\pi^0)$,
we obtain $\calB(\Bz\rightarrow{a_1^+}\rho^-)\calB({a_1^+}\rightarrow (3\pi)^+) < \doublenominalul$. 
This upper limit corresponds to a significant improvement over the previous bound and is compatible
with theoretical expectations~\cite{ref:zpc}. This result is a significant 
improvement in constraining an important \B\ background contribution 
in $\B\to\rho\rho$ decays.
\section{ACKNOWLEDGMENTS}
\label{sec:Acknowledgments}
We are grateful for the excellent luminosity and machine conditions
provided by our \pep2\ colleagues, 
and for the substantial dedicated effort from
the computing organizations that support \babar.
The collaborating institutions wish to thank 
SLAC for its support and kind hospitality. 
This work is supported by
DOE
and NSF (USA),
NSERC (Canada),
IHEP (China),
CEA and
CNRS-IN2P3
(France),
BMBF and DFG
(Germany),
INFN (Italy),
FOM (The Netherlands),
NFR (Norway),
MIST (Russia), and
PPARC (United Kingdom). 
Individuals have received support from CONACyT (Mexico), 
Marie Curie EIF (European Union),
the A.~P.~Sloan Foundation, 
the Research Corporation,
and the Alexander von Humboldt Foundation.



\begin{thebibliography}{99}

\bibitem{CKM}
N.~Cabibbo, \jprl{10}, 531 (1963);
M.~Kobayashi and T.~Maskawa, \progtp {49}, 652 (1973).

\bibitem{charge}
Charge-conjugate transitions are included implicitly unless otherwise stated.

\bibitem{ref:alphareview}
A.\ Bevan, {{Mod.\ Phys.\ Lett.\ A {\bf 21}}}, No.\ 4, 305 (2006).

\bibitem{ref:rhoprhom}
The \babar\ Collaboration, B.\ Aubert {\em et al.},
\jprl{95}, 041805 (2005).

\bibitem{ref:rhoprhoz}
The \babar\ Collaboration, B.\ Aubert {\em et al.},
\jprl{91}, 171802 (2003).

\bibitem{ref:rhozrhoz}
The \babar\ Collaboration, B.\ Aubert {\em et al.},
\jprl{94}, 131801 (2005).

\bibitem{ref:bellerhoprhom}
The Belle Collaboration, A.\ Somov {\em et al.}, 
arXiv:hep-ex/0601024. Submitted to PRL.

\bibitem{ref:bellerhoprhoz}
The Belle Collaboration, J.\ Zhang {\em et al.},
\jprl{91}, 221801 (2003).

\bibitem{ref:argusa1rho}
The ARGUS Collaboration, H.\ Albrecht {\em et al.},
\plb{241}, 278 (1990).

\bibitem{ref:zpc}
M.\ Bauer, B.\ Stech and M.\ Wirbel,
\zpc{34}, 103 (1987). 

\bibitem{ref:babar}
The \babar\ Collaboration, B.\ Aubert {\em et al.},
\nima{479}, 1 (2002).

\bibitem{ref:cleo}
The CLEO Collaboration, D.~M.~Asner {\em et al.},
\jprd{61}, 012002 (1999).  

\bibitem{ref:geant}
The GEANT4 Collaboration, S.\ Agostinelli {\em et al.},
\nima{506}, 250 (2003).

\bibitem{ref:fisher}
R.~A.~Fisher, Annals of Eugenics 7, 179 (1936).

\bibitem{ref:tag}
The \babar\ Collaboration, B.\ Aubert {\em et al.},
\jprl{89}, 281802 (2002).

\bibitem{ref:pdg}
S.\ Eidelman {\em et al.}, 
\plb{592}, 1 (2004). 

\bibitem{ref:keys}
K. S.\ Cranmer, \cpc{136}, 198 (2001).

\bibitem{ref:cb}
The Crystal Ball Collaboration, D.\ Antreasyan {\em et al.},  
Crystal Ball Note, 321 (1983).

\bibitem{ref:argus}
The ARGUS Collaboration, H.\ Albrecht {\em et al.},
\plb{241}, 278 (1990).

\bibitem{ref:splots}
M.\ Pivk and F. R.\ Le Diberder,
\nima{555}, 356 (2005).


\end{thebibliography}
\end{document}